\documentclass[12pt]{article}
\usepackage{amsfonts}
\usepackage{bbm}
\usepackage{graphicx}
\usepackage{sectsty}
\sectionfont{\large}
\subsectionfont{\normalsize}
\usepackage{appendix}
\usepackage{float}
\usepackage{url}
\newcommand{\captionfonts}{\footnotesize}
\makeatletter  
\long\def\@makecaption#1#2{%
  \vskip\abovecaptionskip
  \sbox\@tempboxa{{\captionfonts #1: #2}}%
  \ifdim \wd\@tempboxa >\hsize
    {\captionfonts #1: #2\par}
  \else
    \hbox to\hsize{\hfil\box\@tempboxa\hfil}%
  \fi
  \vskip\belowcaptionskip} 
\makeatother   
\begin{document}
\title{Potentiality States: Quantum versus Classical Emergence}
\author{\normalsize Diederik Aerts and Bart D'Hooghe \\ 
        \small\itshape \vspace{-0.1 cm}
         Center Leo Apostel for Interdisciplinary Studies \\
       \vspace{-0.1 cm} \small\itshape
       Departments of Mathematics and Department of Psychology \\
       \vspace{-0.1 cm} \small\itshape
         Brussels Free University, Brussels, Belgium \\
       \vspace{-0.1 cm} \small
        Emails: \url{diraerts@vub.ac.be, bdhooghe@vub.ac.be} \vspace{0.2 cm} \\
         }
\date{}
\maketitle
\vspace{-0.5 cm}
\begin{abstract}
\noindent 
We identify emergence with the existence of states of potentiality related to relevant physical quantities. We introduce the concept of `potentiality state' operationally and show how it reduces to `superposition state' when standard quantum mechanics can be applied. We consider several examples to illustrate our approach, and define the potentiality states giving rise to emergence in each example. We prove that Bell inequalities are violated by the potentiality states in the examples, which, taking into account Pitowsky's theorem, experimentally indicates the presence of quantum structure in emergence. In the first example emergence arises because of the many ways water can be subdivided into different vessels. In the second example, we put forward a full quantum description of the Liar paradox situation, and identify the potentiality states, which in this case turn out to be superposition states. In the example of the soccer team, we show the difference between classical emergence as stable dynamical pattern and emergence defined by a potentiality state, and show how Bell inequalities can be violated in the case of highly contextual experiments.
\end{abstract}

\section{Introduction}
Many everyday life examples of emergence can be given. Let us consider for instance a set of soccer players. As long as it is just a set of soccer players no emergence takes place. Suppose however that the set of players starts to practice with the aim of forming a soccer team. The co-adaptation that takes place between the different soccer players during their trainings and matches results in the emergence of a soccer team. The soccer team is a new structure that has been formed out of the set of individual soccer players.

In physics emergent phenomena have been studied within the complexity and
chaos approach. An example is given by the B\'{e}nard convection effect.
This effect occurs when a viscous fluid is heated between two planes. In the pre-boiling stage and under suitable conditions, bubbles begin to rise and vortices --- called B\'{e}nard cells --- arise like little cylinders within which the fluid continuously streams up on the outside of the cylinder and back down through the middle of the cylinder. This motion occurs as the fluid heats up and before it starts to boil. The B\'{e}nard cells fit together in a hexagonal lattice of vortices, even without partitions to keep the boundaries between the vortices stable. The microscopical movements of the individual molecules of the fluid result in a macroscopical dynamical pattern of the fluid as a whole with the emergent B\'{e}nard cell structure.

The existing complexity and chaos models are classical physics models. In
this paper we want to show that a classical physics approach has to be
generalized in order to describe emergence in a complete way. The reason is that the emergent structure usually contains states that have a relation of `potential' with respect to relevant observable quantities. We will call such states `potentiality states'. These emergent potentiality states cannot be described in a classical physics approach but need a quantum-like formalism. We know that the appearance of potentiality states is a basic aspect of quantum mechanics. Therefore, in this paper we will demonstrate not only the importance of potentiality states for emergence, but also the advantage of a quantum-like description of emergence versus a classical one. In the examples we will indicate in which way the quantum-like formalism appears in the description. The relevance of quantum aspects in emergence
confirms earlier results in which quantum mechanical aspects in the
macroscopic world are identified in a more general way \cite{aerts82,AeBroGab2000}.

In physics, the state $q(t)$ of a physical entity $S$ at time $t$ represents
the reality of this physical entity at that time $t$.\footnote{In physics `statistical states' are sometimes also just called states. We
however will always refer with the concept state to what is called a `pure
state'.} In the case of classical physics such a state is represented by a
point in phase space, while for quantum physics it is represented by a unit
vector in Hilbert space. In classical physics the state $q(t)$ of a physical
entity $S$ determines the values for all observable quantities at this time $%
t$. Hence it is common in classical physics to characterize the state $q(t)$
by the set of all values of the relevant observable quantities. For example,
if the physical entity is a particle $S(classical \;  particle)$, then the
relevant observable quantities are its position $r(t)$ and momentum $p(t)$
at time $t$. And indeed, for each state $q(t)$ of $S(classical \; particle)$, it
is the case that $r(t)$ and $p(t)$ have definite values, which makes it
possible to represent $q(t)$ with the couple $(r(t),p(t))$ in phase space.
In quantum physics the situation is different. The state $q(t)$ of a quantum
particle $S(quantum \; particle)$ is represented by a unit vector $\psi (r,t)$
(the normalized wave function) in a Hilbert space $L_2(\mathbb{R}^3)$. Again, the relevant observable quantities are the position and the
momentum. However, neither has definite values for the entity $S(quantum \; particle)$ being in a state $\psi (r,t)$. To be more specific,
definite values of position exist only if the wave function is a delta
function, and definite values of momentum if the wave function is a plane
wave.\footnote{If one chooses as the basis of the Hilbert space an orthonormal set of eigenstates corresponding with the position operator, i.e. the state is expressed as a (wave) function $\Psi(x_1,\ldots,x_n;t)$ of variables given by the coordinates $x_i$.} Apart from the fact that in both cases the wave function is not an
element of the Hilbert space $L_2(\mathbb{R}^3)$, and hence should be
considered as limiting cases, they never can occur together. This means that
a quantum particle never has simultaneously a definite position and a
definite momentum, which is referred to as `quantum indeterminism'. It can
also be expressed as follows: for a quantum particle in state $\psi (r,t)$
the values of position and momentum are potential. This means that the
quantum particle has the potential of realizing them, but they are not
actually realized in the state $\psi (r,t)$.

Previously it has been shown that whether a specific physical entity needs a
quantum-like description --- and hence should be called a quantum entity --- or a
classical description, depends on the nature of the entity and the relevant
observable quantities, and not on the fact that it belongs to the microworld
or to the macroworld \cite{aerts82,AeBroGab2000}.

Because of their importance for emergence we will concentrate on the
presence of potentiality states and how without difficulties many situations
can be found in the macroscopic world containing this quantum aspect.
However, they can not be described by classical physical theories which
identify emergent properties of a specific entity with stable dynamical
patterns of the entity. The concept of potentiality state makes it possible
to describe another type of emergence which is present on an ontological
level and not on the dynamic level as in the case of classical emergence.

In standard quantum mechanics a potentiality state related to a specific
observable quantity is a superposition state of the eigenstates of this
observable quantity. Hence, our potentiality states are superposition states
for a situation where standard quantum physics applies. To demonstrate the
more general applicability of the concept of potentiality state, we will
consider situations that have both quantum and classical aspects and hence
are not purely quantum. However for such situations there does not
necessarily exist a Hilbert space describing the set of states, as is the
case in pure quantum situations, meaning that superposition is not well
defined. That is the reason that we use the name potentiality states instead
of superposition states for the states that we consider. In the second
example of this paper (see section 3), the liar paradox, we will see that
the description is fully quantum, such that in this case the potentiality
states reduce to superposition states. In the last section we give the
example of a team of soccer players to illustrate the difference between
emergence due to potentiality states, which is on an ontological level, and
classical emergence, defined by dynamical patterns.

\section{Bell Inequalities and Non-classical Emergence}

In this section we consider several examples and analyze in which way the
concept of potentiality state appears. We also investigate in which way the
presence of potentiality states is linked to the violation of Bell
inequalities.

\subsection{Potentiality States in Connected Vessels of Water}

Suppose that we consider two vessels of which one contains 6 liters of water
and the other one 14 liters of water. We have at our disposal a third vessel
that is empty, but can contain more than 20 liters of water. The entity $%
S(water)$ is the water, and we consider the physical quantity which is the
volume of the water. Clearly this physical quantity has a definite value for
the water that is contained in the two considered vessels, namely 6 liters
and 14 liters. This means that the states of water that we consider in both
vessels are not potentiality states related to the observable quantity which
is the volume.

Suppose now that we take the two vessels and empty them in the third vessel.
This third vessel will then contain 20 liters of water, which means that
also for this new entity of water the physical quantity volume has a
definite value. But by putting the water of the two vessels together, the
new entity of water has lost the old properties of 6 liters and 14 liters as
actual properties. We could however divide the water again and collect 6
liters in one vessel and 14 liters in the other vessel. This means that
potentially the division in 6 liters and 14 liters is still present in the
entity of 20 liters of water. But this new state of 20 liters of water
contains many more possible subdivisions as potentialities. Also 8 liters
and 12 liters, or 11 liters and 9 liters, or 2 liters and 18 liters,$\ldots$ are
all potentialities of subdivisions. In general we can say that $x$ liters
and $y$ liters of water, such that $x+y=20$ form an infinite continuous set
of potential subdivisions. Therefore, from the third vessel, being in the
state such that it contains 20 liters, the two subentities can be derived by
dividing the 20 liters in the appropriate amounts. The state of 20 liters of
water is a potentiality state related to measurements that divide the amount
of water in two amounts. Taking into account the measurement that subdivides
an amount of water in two, the 20 liters of water, that originated by
putting together the original 6 liters and 14 liters has new emergent
properties. These properties are described by the potentiality state, that
allows the water to be subdivided in all these ways.

\begin{figure}[tbh]
\begin{center}
\includegraphics[scale=0.75]{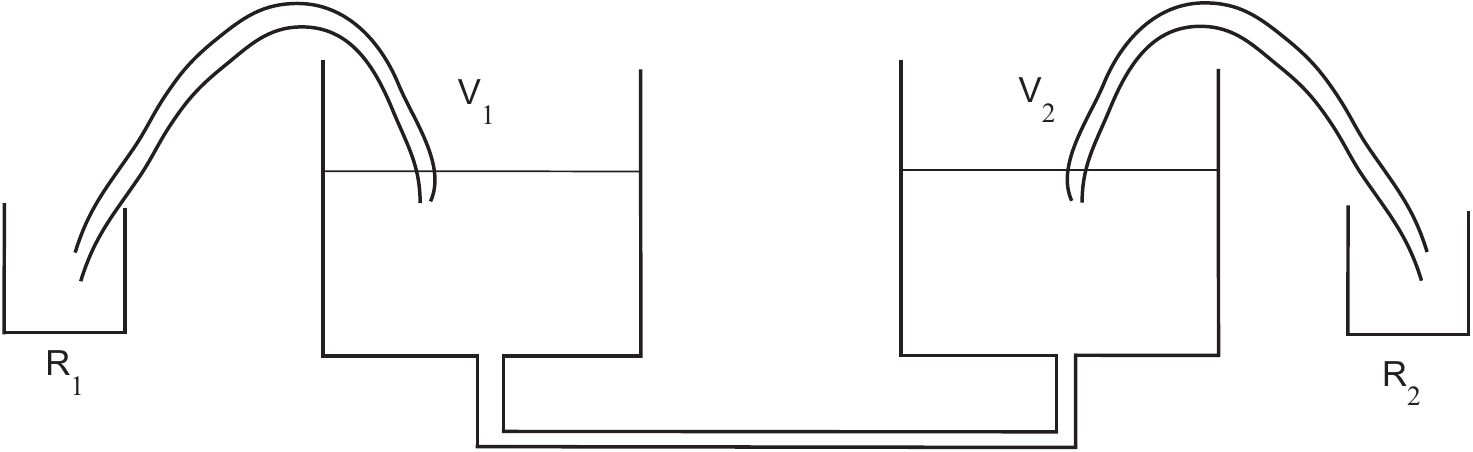}
\end{center}
\caption{The vessels of water example
violating Bell inequalities. The entity $S$ consists of two vessels
containing 20 liters of water that are connected by a tube.}
\label{mqg01}
\end{figure}

If we connect the two original vessels by a tube, we get such a third vessel
(see Figure~\ref{mqg01}). We can show that the new properties that this entity has are
emergent properties related to measurements that divide the water up again.
A criterion that we can use to show the quantum nature of these emergent
properties is the violation of Bell inequalities.
We will recall in the next section Bell inequalities and then elaborate our
connected vessels of water example with the necessary detail such that we
can show that Bell inequalities are violated. The vessels of water example
was introduced in \cite{aerts82} and elaborated in \cite{aerts85a,aerts85b,AeBroGab2000,aerts92}.

\subsection{Bell Inequalities and the Presence of Quantum Structure}

Violation of Bell inequalities related to the presence of potentiality
states can happen as well in the macroworld as in the microworld, depending
on the type of states and observable quantities that are considered. Let us
first recall some of the most relevant historical results related to Bell
inequalities, and show why the violation of Bell inequalities is an
experimental indication for the presence of quantum aspects.

In the seventies, a sequence of experiments was carried out to test for the presence of nonlocality in the microworld described by quantum 
mechanics \cite{clauser76,kas70}, culminating in decisive
experiments by Aspect and his team in Paris \cite{aspect81,aspect82}.
They were inspired by three important theoretical results: the EPR Paradox \cite{epr35}, Bohm's thought experiment \cite{bohm51}, and 
Bell's theorem \cite{bell64}. Einstein, Podolsky and Rosen believed to have shown that quantum mechanics is incomplete, in that there exist elements of reality
that cannot be described by it \cite{epr35,aerts84,aerts2000}. Bohm took
their insight further with a simple example: the `coupled spin-${\frac 12}$
entity' consisting of two particles with spin-${\frac 12}$, of which the
spins are coupled such that the quantum spin vector is a non-product vector
representing a singlet spin state \cite{bohm51}. Bohm's example inspired
Bell to formulate a condition that would test experimentally for
incompleteness, the Bell inequalities \cite{bell64}. Bell's theorem states
that statistical results of experiments performed on a certain physical
entity satisfy his inequalities if and only if the reality in which this
physical entity is embedded is local. Experiments performed to test for the
presence of nonlocality confirmed the results as predicted by quantum
mechanics, such that it is now commonly accepted that the micro-physical
world is incompatible with local realism.

Bell inequalities are defined with the following experimental situation in
mind. We consider a physical entity $S$, and four experiments $e_1$, $e_2$, $%
e_3$, and $e_4$ that can be performed on the physical entity $S$. Each of
the experiments $e_i,i\in \{1,2,3,4\}$ has two possible outcomes,
respectively denoted $o_i(up)$ and $o_i(down)$. Some of the experiments can
be performed together, which in principle leads to `coincidence' experiments 
$e_{ij},i,j\in \{1,2,3,4\}$. For example $e_i$ and $e_j$ together will be
denoted $e_{ij}$. Such a coincidence experiment $e_{ij}$ has four possible
outcomes, namely $(o_i(up),o_j(up))$, $(o_i(up),o_j(down))$, $%
(o_i(down),o_j(up))$ and $(o_i(down),o_j(down))$. Following 
Bell \cite{bell64}, we define the expectation values $\hbox{$\mathbb{E}$}_{ij},i,j\in
\{1,2,3,4\}$ for these coincidence experiments, as 
\begin{equation}
\begin{array}{ll}
\hbox{$\mathbb{E}$}_{ij}= & \left( +1\right) P(o_i(up),o_j(up))+\left( +1\right)
P(o_i(down),o_j(down))+ \\ 
& \left( -1\right) P(o_i(up),o_j(down))+\left( -1\right) P(o_i(down),o_j(up))
\end{array}
\end{equation}
From the assumption that the outcomes are either $+1$ or $-1$, and that the
correlation $\hbox{$\mathbb{E}$}_{ij}$ can be written as an integral over some
hidden variable of a product of the two local outcome assignments, one
derives Bell inequalities:

\begin{equation}
|\hbox{$\mathbb{E}$}_{13}-\hbox{$\mathbb{E}$}_{14}|+|\hbox{$\mathbb{E}$}_{23}+\hbox{$\mathbb{E}$}%
_{24}|\leq 2  \label{bellineq}
\end{equation}
To come to the point where we can use the violation of Bell inequalities as
an experimental indication for the presence of quantum structure, we have to
mention the work of Itamar Pitowsky. Pitowsky proved that if Bell
inequalities are satisfied for a set of probabilities connected to the
outcomes of the considered experiments, there exists a classical
Kolmogorovian probability model. In such model the probability can be
explained as due to a lack of knowledge about the precise state of the
system. If however Bell inequalities are violated, Pitowsky proved that no
such classical Kolmogorovian probability model exists \cite{pit89}. Hence
violation of Bell inequalities shows that the probabilities that are
involved are nonclassical. The only type of nonclassical probabilities that
are well known in nature are the quantum probabilities. The probability
structure that is present in our examples\footnote{%
Except for the liar paradox example, where we derive a pure quantum
description and hence the probability model is quantum.} is nonclassical and
nonquantum, the classical and quantum probabilities being two special cases
of this more general situation.

\subsection{Violation of Bell Inequalities for the Connected Vessels of
Water Entity}

Let us consider again the entity $S(connected \; vessels)$ of two vessels
connected by a tube and containing 20 liters of transparent water (see Figure~\ref{mqg01}). The entity is in an emergent potentiality state $s$ such that the vessel
is placed in the gravitational field of the earth, with its bottom
horizontal. To be able to check for the violation of Bell inequalities
caused by this potentiality state we have to introduce four measurements, of
which some can be performed together.

Let us introduce the experiment $e_1$ that consists of putting a siphon $K_1$
in the vessel of water at the left, taking out water using the siphon, and
collecting this water in a reference vessel $R_1$ placed to the left of the
vessel. If we collect more than 10 liters of water, we call the outcome $%
o_1(up)$, and if we collect less or equal to 10 liters, we call the outcome $%
o_1(down)$. We introduce another experiment $e_2$ that consists of taking
with a little spoon, from the left, a bit of the water, and determining
whether it is transparent. We call the outcome $o_2(up)$ when the water is
transparent and the outcome $o_2(down)$ when it is not. We introduce the
experiment $e_3$ that consists of putting a siphon $K_3$ in the vessel of
water at the right, taking out water using the siphon, and collecting this
water in a reference vessel $R_3$ to the right of the vessel. If we collect
more or equal to 10 liters of water, we call the outcome $o_3(up)$, and if
we collect less than 10 liters, we call the outcome $o_3(down)$. We also
introduce the experiment $e_4$ which is analogous to experiment $e_2$,
except that we perform it to the right of the vessel (see Figure~\ref{mqg01}).




The experiment $e_1$ can be performed together with experiments $e_3$ and $%
e_4$, and we denote the coincidence experiments $e_{13}$ and $e_{14}$. Also,
experiment $e_2$ can be performed together with experiments $e_3$ and $e_4$,
and we denote the coincidence experiments $e_{23}$ and $e_{24}$. For the
vessel in state $s$, the coincidence experiment $e_{13}$ always gives one of
the outcomes $(o_1(up),o_3(down))$ or $(o_1(down),o_3(up))$, since more than
10 liters of water can never come out of the vessel at both sides. This
shows that $\hbox{$\mathbb{E}$}_{13}=-1$. The coincidence experiment $e_{14}$
always gives the outcome $(o_1(up),o_4(up))$ which shows that $\hbox{$\mathbb{E}$}%
_{14}=+1$, and the coincidence experiment $e_{23}$ always gives the outcome $%
(o_2(up),o_3(up))$ which shows that $\hbox{$\mathbb{E}$}_{23}=+1$. Clearly
experiment $e_{24}$ always gives the outcome $(o_2(up),o_4(up))$ which shows
that $\hbox{$\mathbb{E}$}_{24}=+1$. Let us now calculate the terms of Bell
inequalities, 
\begin{equation}
\begin{array}{ll}
|\hbox{$\mathbb{E}$}_{13}-\hbox{$\mathbb{E}$}_{14}|+|\hbox{$\mathbb{E}$}_{23}+\hbox{$\mathbb{E}$}%
_{24}| & =|-1-1|+|+1+1| \\ 
& =+2+2 \\ 
& =+4
\end{array}
\end{equation}
This shows that Bell inequalities are violated. The state $s$, which is a
potentiality state related to the measurements that divide up the 20 liters
of water in the two reference vessels, is at the origin of this violation.

\section{States of Potentiality and the Liar Paradox}

\subsection{The Cognitive Entity of the Liar Paradox}

Another example of an entity for which a description with potentiality
states is useful, is given by the double liar paradox. In fact, it will turn
out that the double liar paradox entity $S(double \; liar)$ can be given a pure
quantum mechanical description. As we shall show below, the potentiality
state of this entity will be given by a superposition state of the states of
the sub-entities which compose the double liar entity. Before discussing the
double liar paradox, let us first consider some non-paradoxical situations:
\begin{itemize}
\item[(S 0.1)]  The sum of two and two is four.
\item[(S 0.2)] This page contains 1764 letters.
\item[(S 0.3)] The square root of 1764 is 42.
\item[(S 0.4)] 6 times 9 yields 42.
\end{itemize}

\noindent
Obviously, the truth-value of each sentence can be easily determined. The
fact that $2+2=4$ can also be expressed by the set of following two
sentences:
\begin{itemize}
\item[(S 1.1)] Sentence (S 1.2) is true.
\item[(S 1.2)] The sum of two and two is four.
\end{itemize}

\noindent
The truth behavior of the entity, consisting of the set of sentences (S 1.1)
and (S 1.2), can also be determined rather easily. The truth values of the
sentences are coupled, but we do not encounter any paradoxical situations.
This happens if both sentences refer to the truth value of each other. In
such case, paradoxical situations arise. Let us now consider the double liar
paradox which can be presented in the following way:

\begin{center}
\bigskip \textsl{`Double Liar'}

\medskip (1) \ \ \ sentence (2) is false

\medskip (2) \ \ \ sentence (1) is true
\end{center}

\noindent
Let us describe a typical cognitive interaction that one goes through with
these liar paradox sentences. Suppose we hypothesize that sentence (1) is
true. We go to sentence (1) and read what is written there. It is written
`sentence (2) is false'. From our hypothesis we can infer that sentence (2)
is false. Let us go, using this knowledge about the status of sentence (2),
to sentence (2). There is written `sentence (1) is true'. From the knowledge
that sentence (2) is false, we can infer that sentence (1) is false. This
means that from the hypothesis that sentence (1) is true we derive that
sentence (1) is false, which is a contradiction. Similarly, starting from
the hypothesis that sentence (1) is false one also obtains a contradiction.

In the examples presented in the previous section the presence of
potentiality states gives rise to the violation of Bell inequalities, which
implies that there is no classical, i.e., Kolmogorovian, representation
possible \cite{pit89}. Therefore in general, potentiality states are
nonclassical states. In \cite{AeBroSme99,AeBroSme2000} the entity which is
the liar sentence is studied in a similar perspective. In the description of
the liar entity we interpret the interaction that a person has with this
entity, as described in some detail in the forgoing sections, as a
measurement. It is with the introduction of the concept of entity and of
measurement (or observable quantity) in this operational way, that we can
find out what the nature is of this liar entity. It turns out that the liar
entity can be described using the formalism of standard quantum mechanics in
a complex Hilbert space. Moreover, the self-referential circularity --- more
precisely, the truth-value dynamics --- of the liar paradox can be described
by the Schr\"{o}dinger equation.

\subsection{Quantum Representation for the Potentiality State of the Liar Paradox}
A potentiality state is necessary to represent the state of the entity
defined by the liar paradox since its state is not an eigenstate of `truth'
(i.e., the sentence(s) are true) neither an eigenstate of `falsehood' (i.e.,
the sentences are not true). Instead, the state of the liar entity can be
regarded as a potentiality state related to the measurements introduced by
the persons that interact cognitively with the liar entity. Since we find a
full quantum description in this case, it follows that the potentiality
state is a superposition state of both eigenstates, just as the singlet
state in the case of two correlated spin-$\frac 12$ entities.

Let us discuss the Double Liar sentences in more detail by considering
following three situations: 
\[
\mathrm{A}\ \ \left\{ 
\begin{array}{ll}
\mathrm{(1)\ } & \mathrm{sentence\ (2)\ is\ false} \\ 
\mathrm{(2)\ } & \mathrm{sentence\ (1)\ is\ true}
\end{array}
\right. 
\]

\[
\mathrm{B}\ \ \left\{ 
\begin{array}{ll}
\mathrm{(1)\ } & \mathrm{sentence\ (2)\ is\ true} \\ 
\mathrm{(2)\ } & \mathrm{sentence\ (1)\ is\ true}
\end{array}
\right. 
\]

\[
\mathrm{C}\ \ \left\{ 
\begin{array}{ll}
\mathrm{(1)\ } & \mathrm{sentence\ (2)\ is\ false} \\ 
\mathrm{(2)\ } & \mathrm{sentence\ (1)\ is\ false}
\end{array}
\right. 
\]
Since truth value of a sentence is binary: either it is true or it is false,
we can associate the outcomes of a dichotomic observable (e.g., the outcomes
for a spin-$\frac 12$ outcome are either `spin up' or `spin down') with
truth and falsehood. Let us make the convention that the `spin up' state $%
\left( 
\begin{array}{c}
1 \\ 
0
\end{array}
\right) $ corresponds with truth of a sentence and `spin down' state $\left( 
\begin{array}{c}
0 \\ 
1
\end{array}
\right) $ with the falsehood of the respective sentence. It turns out that
then for the paradoxes of type B and C the sentences can be represented by
coupled $\Bbb{C}^2$ vectors. Indeed, since for each measurement the truth
values of the two sentences are coupled (B1 true implies B2 true and B1
false implies B2 false, C1 true implies C2 false and vice versa), the
dimensionality of {$\Bbb{C}$}$^2\otimes $\thinspace {$\Bbb{C}$}$^2$ is
sufficient to represent these entities. Formally, the corresponding quantum
mechanical representation would be by a `singlet state' and `triplet state'
respectively. In the singlet state the two spin-1/2 particles are
anti-alined and in an anti-symmetrical state (entity C), while in a triplet
state the two spin-1/2 particles are alined and in a symmetrical state
(entity B). In the specific case of (C), and taking into account the
anti-symmetric spin analog, the state of the entity $\Psi $ is given by: 
\[
\frac 1{\sqrt{2}}\left\{ \left( 
\begin{array}{c}
1 \\ 
0
\end{array}
\right) \otimes \left( 
\begin{array}{c}
0 \\ 
1
\end{array}
\right) -\left( 
\begin{array}{c}
0 \\ 
1
\end{array}
\right) \otimes \left( 
\begin{array}{c}
1 \\ 
0
\end{array}
\right) \right\} 
\]
Equivalently, the liar entity can be represented by other linear
combinations of $\left( 
\begin{array}{c}
1 \\ 
0
\end{array}
\right) \otimes \left( 
\begin{array}{c}
0 \\ 
1
\end{array}
\right) $ and $\left( 
\begin{array}{c}
0 \\ 
1
\end{array}
\right) \otimes \left( 
\begin{array}{c}
1 \\ 
0
\end{array}
\right) ,$ provided the coefficients have equal amplitude and the squared
amplitudes add to one (such that total probability of finding the entity in
one of the two possible states after a measurement equals one). In a similar
manner the liar paradox in the `symmetrical' case~(B) can be represented by
the triplet state: 
\[
\frac 1{\sqrt{2}}\left\{ \left( 
\begin{array}{c}
1 \\ 
0
\end{array}
\right) \otimes \left( 
\begin{array}{c}
1 \\ 
0
\end{array}
\right) +\left( 
\begin{array}{c}
0 \\ 
1
\end{array}
\right) \otimes \left( 
\begin{array}{c}
0 \\ 
1
\end{array}
\right) \right\} 
\]

\noindent The projection operators which make sentence 1, respectively
sentence 2, true are: 
\[
P_{1,true}=\left( 
\begin{array}{cc}
1 & 0 \\ 
0 & 0
\end{array}
\right) \otimes \mathbb{1}_2\ \ \ \ \ \ P_{2,true}=\mathbb{1}_1\otimes
\left( 
\begin{array}{cc}
1 & 0 \\ 
0 & 0
\end{array}
\right) 
\]

\noindent The projection operators that make the sentences false are
obtained by switching the elements $1$ and $0$ on the diagonal of the
matrix. These four projection operators represent all possible `logical'
interactions (measurement interactions) between the cognitive observer and
the liar entity. During the measurement process carried out on the entities
(B) and (C), the observer attributes truth-value to the sentences in a
repetitive manner: in case of entity (B) it is a repetition of true-states
(resp.~false-states) depending on whether an initial true (resp.~false)
state was presupposed. In the case of entity (C) it will be an alternation
between true-states and false states, no matter which state was presupposed.

Finally, let us indicate the main points necessary in the derivation of a
full description of the original double liar paradox, i.e., case (A), and
see what pattern of truth assignments follows during the measurement (we
refer to \cite{AeBroSme99} for a more detailed discussion of the derivation
of the results). While for cases (B) and (C) the dimensionality of the
coupled Hilbert space {$\Bbb{C}$}$^2\otimes $\thinspace {$\Bbb{C}$}$^2$ is
sufficient, a space of higher dimension has to be used for case (A). This is
due to the fact that no initial state can be found in the restricted space {$%
\Bbb{C}$}$^2\otimes $\thinspace {$\Bbb{C}$}$^2$, such that application of
the four true--false projection operators results in four orthogonal states
respectively representing the four truth--falsehood states. The existence of
such a superposition state --- with equal amplitudes of its components ---
is required to describe the state of the entity before and after the
measurement process. Since the truth-values of the two sentences in the
paradox are not anymore coupled like it was the case for (B) and (C), the
dynamical pattern of truth assignment by the observer is not anymore a
two-step process but a four-step process. Therefore, the entity should be
described in a 4 dimensional Hilbert space for each sentence. The Hilbert
space needed to describe the Double Liar~(A) is therefore 
$\Bbb{C}^4 \otimes \Bbb{C}^4$.

The initial un-measured superposition state --- $\Psi _0$ --- of the Double
Liar~(A) is given by following superposition of the four true--false states:

{\small 
\[
\frac 12\left\{ 
\begin{array}{c}
\left( 
\begin{array}{c}
0 \\ 
0 \\ 
1 \\ 
0
\end{array}
\right) \otimes \left( 
\begin{array}{c}
0 \\ 
1 \\ 
0 \\ 
0
\end{array}
\right) +\left( 
\begin{array}{c}
0 \\ 
1 \\ 
0 \\ 
0
\end{array}
\right) \otimes \left( 
\begin{array}{c}
0 \\ 
0 \\ 
0 \\ 
1
\end{array}
\right) 
+\left( 
\begin{array}{c}
0 \\ 
0 \\ 
0 \\ 
1
\end{array}
\right) \otimes \left( 
\begin{array}{c}
1 \\ 
0 \\ 
0 \\ 
0
\end{array}
\right) +\left( 
\begin{array}{c}
1 \\ 
0 \\ 
0 \\ 
0
\end{array}
\right) \otimes \left( 
\begin{array}{c}
0 \\ 
0 \\ 
1 \\ 
0
\end{array}
\right)
\end{array}
\right\} 
\]
} Each term in this superposition state is the consecutive state which is
reached in the course of time, when the paradox is reasoned through. The
truth--falsehood values attributed to these states, refer to the chosen
measurement projectors.

\noindent Making a sentence true or false in the act of measurement, is
described by the appropriate projection operators in $\Bbb{C}^4 \otimes \Bbb{C}^4$. In the case we make sentence 1 (resp. sentence
2) true we get: 
\begin{eqnarray*}
P_{1,true} &=&\left( 
\begin{array}{cccc}
0 & 0 & 0 & 0 \\ 
0 & 0 & 0 & 0 \\ 
0 & 0 & 1 & 0 \\ 
0 & 0 & 0 & 0
\end{array}
\right) \otimes \mathbb{1}_2 \\
P_{2,true} &=&\mathbb{1}_1\otimes \left( 
\begin{array}{cccc}
0 & 0 & 0 & 0 \\ 
0 & 0 & 0 & 0 \\ 
0 & 0 & 1 & 0 \\ 
0 & 0 & 0 & 0
\end{array}
\right)
\end{eqnarray*}
The projectors for the false-states are constructed by placing the $1$ on
the final diagonal place: 
\begin{eqnarray*}
P_{1,false} &=&\left( 
\begin{array}{cccc}
0 & 0 & 0 & 0 \\ 
0 & 0 & 0 & 0 \\ 
0 & 0 & 0 & 0 \\ 
0 & 0 & 0 & 1
\end{array}
\right) \otimes \mathbb{1}_2 \\
P_{2,false} &=&\mathbb{1}_1\otimes \left( 
\begin{array}{cccc}
0 & 0 & 0 & 0 \\ 
0 & 0 & 0 & 0 \\ 
0 & 0 & 0 & 0 \\ 
0 & 0 & 0 & 1
\end{array}
\right)
\end{eqnarray*}

\noindent Starting by making one of the sentences of (A) either true or
false, by logical inference, the four consecutive states are run through
repeatedly. In \cite{AeBroSme99} a continuous parameter $t$ was introduced
to reflect `reasoning time' of the cognitive observer. This allowed to give
a time-ordered description of the cyclic change of state present in the
measurement process of the liar paradox. A Hamiltonian $H$ can be
constructed such that the unitary evolution operator $U(t)$ --- with $%
U(t)=e^{-iHt}$ --- describes this cyclic change.

\noindent The dynamical picture of the Double Liar cognitive entity (A) is
therefore as follows: when submitted to measurement, the entity starts its
truth--falsehood cycle; when left un-measured the entity remains in the
potentiality state. This follows immediately from the fact that the initial
state $\Psi _0$ is left unchanged by the dynamical evolution $U(t)$: 
\[
\Psi _0(t)=\Psi _0 
\]
because $\Psi _0$ is a time invariant, as an eigenstate of the Hamiltonian $%
H $. Because of this time-independence, the state $\Psi _0$ describes the
cognitive entity of the liar paradox A, regardless of the observer. The
highly contextual nature of the Double Liar (A) --- its unavoidable dynamics
induced by the measurement process --- implies that intrinsically it can not
expose its complete nature, analogous to the quantum entities of the
micro-physical world.

\section{Potentiality States versus Classical Emergence}

\subsection{Potentiality State of a Soccer Team versus its Stable Classical Dynamical Pattern}

Classical emergent properties of a system are defined by dynamical patterns
related to its subsystems. Each of these subsystems can be described in a
state space, and therefore the emergent properties which are identified with
the collection of the subsystems, can be represented in essentially the same
space. When potentiality states are involved, the emergent entity needs to
be described in a higher dimensional state space, as the example of the liar
paradox shows where the potentiality state is a superposition of tensor
products of vectors representing the individual sentences. Crucial for the
existence of a potentiality state is the contextuality in the measurement
process, which causes a `collapse' of the potentiality state into one of the
mutual exclusive possible eigenstates. Let us now clarify the main
differences between a classical emergent pattern due to dynamical evolution
of a complex system, and the new kind of emergence related to the existence
of potentiality states, by the concrete example of a soccer team.

A simple example of an entity represented with a potentiality state is given
by a soccer team, e.g. the team of R.F.C. Anderlecht of Belgium
(abbreviated: the A-team), consisting of eleven soccer players. The
classical emergent description of such soccer team would involve analyzing
the team in terms of the individual motions of the players and would reflect
the dynamical pattern which is present during a game: e.g., whether the team
as a whole is attacking or defending at one moment in time. The properties
of the soccer team are determined by the capabilities of each player (how
fast each player can run, how good they can pass the ball etc.), and on the
correlation which exists between the players, i.e., how the members of the
team play together as a team. Hence, the whole soccer team can be identified
with the eleven members of the team and the team can be represented in terms
of the descriptions of the individual players: indeed, once the movement of
each player is known, the movement of the team as a whole is known too. This
is what we would call the team as a `classical' emergent entity. Notice,
that when the eleven players have left the field after the game, the entity
of the team stops to exist. Indeed, the movements of the eleven players
become incoherent after the game: each player leaves to his own home, and
until they have to play the next match, they have stopped to be, in
classical emergent terms, part of the larger entity `a soccer team'. For
instance, one of the players of the A-team could be a foreigner who has to
play a game for his native country before the next match of the A-team.
Obviously, while he's playing the match for his native country, he cannot be
playing a match for the A-team. Therefore, while he's playing for his home
country, the classical emergent entity `the A-team' defined by the eleven
soccer players does not exhibit the typical dynamical patterns of a soccer
team playing a match, and therefore the entity does not exist at that moment
in time.

If we elaborate this example in terms of the potentiality state concept, we
obtain the following. The state of the entity `the A-team' is defined by the
possible instances of the team as a collection of the individual players. By
this we mean that when the eleven players are on the (same) soccer field
playing a game for the A-team, they are in fact actualizing a potential
game, simply by the act of playing the game. The emergent concept of `the
A-team' is more than just the collection of the movements of the eleven
players. Even when the eleven players are not actually playing a game, they
are still a `potential team'. The instance of the team playing a game is
created at the moment the A-team is actually playing a game. To make this
more clear, let us consider the following situation. The A-team has
qualified for the final of the Belgian soccer cup in which they will have to
play against the winner of the other semi-final match Bruges-Ghent which is
played a day later than the A-team semi-final. Then `the A-team playing the
cup final' can be considered as an emergent entity, defined by the eleven
players who are preparing themselves for the final. A day later the match
Bruges-Ghent will be actually played and the adversary of the A-team in the
final will be known. At the moment the final is played, the instance `the
A-team is playing' is activated, and potentiality of `the A-team playing a
game' collapses into one of two mutually exclusive possibilities, i.e., a
final against Bruges or a final against Ghent. Notice that even when the
A-team is not actually playing, the emergent concept defined by the
potentiality of letting the eleven players play a game is still present. As
such, `the A-team playing the final' can be viewed as a potentiality state
of two possible instances: one concerns the final against the (let's say
defending) team of Bruges, and one against the (let's say attacking) team of
Ghent. Depending on which team reaches the final, the A-team will behave
differently because they have to choose between different tactics.
Nevertheless, before the score of the semi-final Bruges-Ghent is determined,
potentially both tactics could be followed. It is not just the state of the
A-team which decides how the final will look like, also the adversary will
be decisive, and the result of the game Bruges-Ghent is not influenced by
the state of the A-team.

The differences between classical concept of emergence and the emergent
behavior which is established by the presence of potentiality states,
resemble the differences in dynamical evolution of a classical system versus
a quantum system. The evolution of a classical system is continuous, and the
act of measurement does not influence the results, the measurements are
purely non-perturbative observations. As such, the dynamical pattern of the
system as a whole is identified by the dynamical evolution of the
sub-systems. In other words, the state of the compound system evolves
continuously in time.

The dynamical evolution for a quantum particle, upon which no measurements
are carried out, is given by the Schr\"{o}dinger equation, which also
describes a continuous evolution of the state of the entity. However,
standard quantum theory predicts a non-continuous evolution during the
measurement process, such that the state of the system changes
instantaneously towards one of the possible eigenstates corresponding with
the observable. Similarly, the potentiality of the eleven players of the
A-team to play a game against Bruges or a game against Ghent defines an
emergent entity. Before the result of Bruges-Ghent is known, potentially the
A-team can play a game against Ghent, but also potentially against Bruges.
Nevertheless, only one of these possible finals can actually be played.
Which final will be played, does not depend on the A-team alone, but also on
the adversary who manages to qualify: Bruges or Ghent. This dependence on
the context (i.e., the semi-final between Bruges and Ghent) causes the
non-continuous evolution of `the A-team playing the final' from a
potentiality of both games towards one of the specific instances. As such,
one can regard the concept of potentiality state as a possible way to
generalize the concept of classical emergence to cases where also
contextuality is important. Eventually, this should lead to a general theory
capable of describing on the one hand the continuous evolution of a system
in the absence of contextual interaction and on the other hand the
discontinuous evolution of the system during the interaction process with
the environment during which in a highly contextual way one of the
possibilities present in the potentiality state of the system is effectively
actualized.

\subsection{Violation of Bell Inequalities for the Soccer Teams}

Let us now show that also in the case of a soccer team we can define a set
of experiments for which Bell inequalities are violated, indicating the
presence of a potentiality state. Depending on the context only one of
mutual exclusive potential outcomes will be actually realized during the
experiment. The entity that we consider is the set of 22 soccer players of
two teams playing a cup final.

The first experiment $e_1$ consists in letting someone give money to a
player of team A such that he will cause his team to lose the cup final. If
his team actually loses the cup final, the experiment $e_1$ is said to yield
the outcome $o_1(up)$, if team A wins the final, the experimental outcome is 
$o_1(down)$. Taking into account that the final is played until one of the
two teams wins (e.g., in the case of a draw the final could be decided with
a number of penalties), the experiment will always give one of these two
possible outcomes. The second experiment $e_2$ consists in looking whether
the referee gives a player of team A a yellow card or not. If he actually
gives a yellow card to at least one player of team A, the outcome is $o_2(up)
$, and $o_2(down)$ otherwise. The experiment $e_3$ consists in letting
someone give money to a player of team B such that he will cause his team to
lose the cup final. If team B actually loses the cup final, the experiment $%
e_3$ is said to yield the outcome $o_3(up)$, if team B wins the final, the
experimental outcome is $o_3(down)$. The fourth and last experiment $e_4$
consists in looking whether the referee gives a player of team B a yellow
card. If he actually gives a yellow card to at least one player of team B,
the outcome is $o_4(up)$, and $o_4(down)$ otherwise. Finally, we assume that
the referee has a bad character such that during the final he will
definitely give a yellow card to at least one player of team A and one
player of team B. Let us now look at the coincidence experiments $e_{13}$, $%
e_{14}$, $e_{23}$ and $e_{24}$. The experiment $e_{14}$ consists in giving a
player of team A money with the aim of letting team A lose the final, and
looking whether a player of team B has received a yellow card. Of course,
even if the other players of team A are playing very good, the bribed player
can make intentionally mistakes like own-goals etc.~such that even in that
case team A loses the final. Therefore, the experiment $e_{14}$ gives the
outcome $(o_1(up),o_4(up))$ and the expectation value $\hbox{$\mathbb{E}$}%
_{14}=+1.$ Similarly, we obtain that $\hbox{$\mathbb{E}$}_{23}=+1.$ Also, due to
our assumption about the bad-character referee, we can deduce that the
coincidence experiment $e_{24}$ yields always the outcome $(o_2(up),o_4(up))$
such that the expectation value is given by $\hbox{$\mathbb{E}$}_{24}=+1.$
Finally, let us look at the coincidence experiment $e_{13}.$ In this case,
both a player of team A and a player of team B will receive money to let
their respective team lose the cup final. However, only one of the two
teams can actually lose the final. Therefore, the coincidence experiment $%
e_{13}$ can only yield the outcome $(o_1(up),o_3(down))$ or $%
(o_1(down),o_3(up)),$ resulting in an expectation value $\hbox{$\mathbb{E}$}%
_{13}=-1.$ Let us now calculate the terms of Bell inequalities,

\[
\begin{array}{ll}
|\hbox{$\mathbb{E}$}_{13}-\hbox{$\mathbb{E}$}_{14}|+|\hbox{$\mathbb{E}$}_{23}+\hbox{$\mathbb{E}$}%
_{24}| & =|-1-1|+|+1+1| \\ 
& =+2+2 \\ 
& =+4
\end{array}
\]
and it follows that Bell inequalities are violated. This violation is due to
the explicit contextuality of the experimental outcomes. To make this more
clear, we could also specify in the definition of the experiments $e_1$ and $%
e_3$ the amount of money which is given to the players. For instance, if we
specify that in experiment $e_1$ the bribed player is very poor, and the
amount of money is to be one billion dollars, then he will probably do
everything possible in order to let his team lose the final. If on the
other hand, the player in experiment $e_3$ is already rich, and the amount
of money is only 100.000 dollars, he will probably not do such extreme
things like making a lot of own-goals etc., which the other, poor player
will probably do to earn his billion dollars. This shows that also in the
case of deterministic coincidence experiments ($e_{13}$ always yielding the
outcome $(o_1(up),o_3(down))$) one can violate Bell-inequalities.

The reason why Bell-inequalities are also violated in the deterministic case
is situated in the fact that only one of two possible but mutually exclusive
situations can occur (i.e., only one of the two teams can lose the cup
final), and that before the experiments are actually performed both
experimental outcomes are possible. It is the contextual nature of the
experiments which defines which of the possible experimental results will
actually occur. This shows that the emergent phenomenon of two teams playing
the cup final should be described with a potentiality state, such that in
one of the possible cases team A loses the final, and in the other team B
loses the final. Depending on which experiment is actually performed (which
player is given what amount of money), the actually played final will be
different. Therefore, the final cannot be regarded only as the stable
dynamical pattern exhibited by the two teams of soccer players playing a
game.

\section{Conclusions}

The above examples --- the `vessels of water', `liar paradox' and `the soccer
team' --- illustrate the various ways in which potentiality states can be
identified in reality. The emergent properties which the potentiality states
define, depend on the particular nature of the entity. For the vessels of
water example, the appearance of potentiality states indicates the presence
of a particular quantum aspect, which can be demonstrated by the violation
of Bell inequalities. For the liar paradox example we have worked out a full
and detailed quantum description which shows the pure quantum mechanical
nature of the example. The potentiality states in this case are
superposition states. We mention that the example of the liar paradox shows
that the reality of conceptual space is quite different from great part of
the macroscopic world. It seems that it contains `Hilbert space-like'
features, which makes it quantum-like. We have shown on the example of a
soccer team that the emergence due to potentiality states has a quite
different status than the one of the emergent dynamical patterns that are
identified in classical physics. The emergent potentiality states are
ontological states within the formalism, and not connected to dynamical
patterns alone. Also in the case of the soccer teams playing a match, one
can define experiments such that Bell inequalities are violated, indicating
the contextual nature of the defined experiments for these entities.

\bigskip
\noindent
{\bf Acknowledgments}

\bigskip
\noindent
The authors would like to acknowledge the support by the Fund for Scientific Research--Flanders (Belgium)(F.W.O.--Vlaanderen).

\end{document}